\documentclass[useAMS,usenatbib,usegraphicx]{mn2e}

\usepackage{color}
 \usepackage{hyperref}
 \providecommand{\adsurl}[1]{\href{#1}{ADS}}
  
\title[Imprints of the anisotropic inflation on the cosmic microwave background]{Imprints of the anisotropic inflation on the cosmic microwave background}
\author 
[M. Watanabe, S. Kanno, J. Soda]
{ Masa-aki Watanabe$^{1}$
, Sugumi Kanno$^{2}$, and Jiro Soda$^{1}$\\
$^{1}$Department of Physics, Kyoto University, Kitashirakawa-oiwake-cho, Sakyo, Kyoto, 606-8502, Japan\\
$^{2}$Department of Physics and Astronomy, Tufts University, Robinson Hall, 212 College Avenue,Medford, MA, 02155, USA
}
\begin{document}



\maketitle

\label{firstpage}

\begin{abstract}
We study the imprints of anisotropic inflation
on the CMB temperature fluctuations and polarizations.
The statistical anisotropy stems not only from the direction dependence of curvature and tensor perturbations, but also from the cross correlation between curvature and 
tensor perturbations, and the linear polarization of tensor perturbations. We show that
off-diagonal $TB$ and $EB$ spectra as well as on- and off-diagonal $TT, EE, BB, TE$ spectra are induced from anisotropic inflation. We emphasize that 
the off-diagonal spectra induced by the cross correlation could be
a characteristic signature of anisotropic inflation.
\end{abstract}


\section{Introduction}
Precise observations of the cosmic microwave background radiation (CMB) enable us to test the fundamental predictions of inflation on primordial fluctuations such as scale independence and Gaussianity. The statistical isotropy has been a robust 
prediction protected by the cosmic no-hair conjecture
 which claims that the inflation washes out classical anisotropy.
Recently, however, its apparent violation has been reported and its origins including systematic effects have been widely discussed.\citep{ArmendarizPicon:2005jh, Pullen:2007tu,Dvorkin:2007jp,Groeneboom:2008fz,Hanson:2009gu,ArmendarizPicon:2008yr,Groeneboom:2009cb,Hanson:2010gu,Pontzen:2010yw,Bennett:2010jb}

The statistical anisotropy in primordial power spectrum of curvature fluctuations
has been discussed so far. While, tensor perturbations (primordial gravitational waves) have not been taken into account. However, since the statistical anisotropy could be a relic of the violation of the rotational symmetry in the early universe, we can naturally expect that there exists cross correlation between curvature and tensor perturbations since they used to interact with each other at that time. Its effects on the CMB would be relevant.
In fact, an anisotropic inflation proposed by us exhibits such features 
\citep{Watanabe:2009ct}. 

We can test the statistical anisotropy in tensor perturbations
with CMB $B$-mode polarization and its cross correlation with temperature fluctuations and $E$-mode polarization. In the conventional cosmology, $B$-mode polarization is supposed to have no two-point cross correlation with temperature or $E$-mode. This results from the two assumptions: statistical isotropy and parity invariance. The statistical isotropy obliges the correlations to be diagonal $\langle a_{lm}^{X}a_{l'm'}^{X'} \rangle=C_{l}^{XX'}\delta _{ll'} \delta _{mm'}$, while $a_{lm}^{T,E}$ and $a_{lm}^{B}$ have parity $(-1)^l$ and $(-1)^{l+1}$, respectively. Hence, correlations $C_l^{TB,EB}$ have odd parity and vanish due to the parity invariance. In the presence of the statistical anisotropy, off-diagonal ($l\neq l'$) correlations arise and therefore $TB,EB$ correlations can appear even if the parity is conserved. These signals may be useful to distinguish the origins of apparent violation of statistical isotropy.

In this paper, we discuss the statistical anisotropy from anisotropic inflation
including tensor perturbations and show how they are imprinted in the two-point correlations of the CMB temperature fluctuations and polarizations. Then we 
compare signals of the primordial anisotropy induced by tensor perturbations with that induced purely by scalar perturbations, and see what we can expect as 
a signal peculiar to anisotropic inflation.
In the next section, we define the statistical anisotropy which we deal with. In section 3, angular power spectra are evaluated. 
The final section is devoted to the discussion. 

\section{primordial statistical anisotropy}

In the anisotropic inflation model, we have four kinds of anisotropy: (i) direction dependence in primordial power spectrum of scalar (curvature) perturbations, (ii) that in tensor perturbations, (iii) cross correlation between curvature perturbations and a linear polarization mode of tensor perturbations, and (iv) linear polarization of tensor perturbations.

It is convenient to express the primordial power spectra in the following way:
\begin{equation}
\langle R^s({\bf k}) R^{s'}({\bf k'})^* \rangle = P^{ss'}({\bf k}) \delta ^3 ({\bf k}-{\bf k'}), \label{eq:power}
\end{equation}
where primordial scalar perturbations is represented by $R^0$ and right- and left-handed circular polarizations of tensor perturbations are given by $R^{+2}$ and $R^{-2}$ respectively, while the delta function results from an assumption of statistical homogeneity (translational invariance).
In the conventional isotropic inflation, where rotational invariance holds, $P^{ss'}({\bf k})$ are proportional to $\delta _{ss'}$ and functions dependent only on $|s|$ and $|{\bf k}|$. 
When we consider the cross correlation (iii) and  the linear polarization (iv),
 we need to take into account off-diagonal ($s\neq s'$) components. 

Note that, unlike diagonal ones, these components change their values with the rotation of the polarization bases, hence the bases have to be specified. The simplest choice is to make use of spherical coordinates with a certain fixed preferred direction, i.e. $\sqrt{2}e^{\pm 2}_{ij}=e_{ij}^{\rm d}\pm ie_{ij}^{\times},\ \sqrt{2}e_{ij}^{{\rm d}}={\bf e}_i^{\theta} {\bf e}_j^{\theta} - {\bf e}_i^{\phi} {\bf e}_j^{\phi},\ \sqrt{2}e_{ij}^{\times}={\bf e}_i^{\theta} {\bf e}_j^{\phi} + {\bf e}_i^{\phi} {\bf e}_j^{\theta},$ 
here $\theta$ and $\phi$ are polar and azimuthal angles. This convention is adopted throughout this paper. The power spectra in linear polarization modes $\sqrt{2}R^{{\rm d}}=R^{+2}+R^{-2},\ \sqrt{2}R^{\times}=i(R^{+2}-R^{-2})$ are similarly defined by $\langle R^{\alpha}({\bf k}) R^{\beta *}({\bf k'}) \rangle = P^{\alpha \beta}({\bf k}) \delta ^3({\bf k}-{\bf k'})$, where $\alpha,\ \beta$ denote $0,{\rm d}$ or $\times$.
Then, the power spectra of helicity bases are expressed by those of linear bases as 
\begin{eqnarray}
P^{0\pm 2}&=&(P^{0{\rm d}}\pm iP^{0\times})/\sqrt{2},\ P^{\pm 2 0}=(P^{{\rm d}0}\mp i P^{\times 0})/\sqrt{2}, \nonumber\\
P^{+2+2} &=& P^{-2-2} = (P^{\rm dd}+P^{\times \times})/2\equiv P_{t}^{\rm unp}\nonumber\\
P^{+2-2} &=& P^{-2+2}=(P^{\rm dd}-P^{\times \times})/2\equiv P_{t}^{\rm pol}, \label{eq:relation}
\end{eqnarray}
We hereafter neglect circular polarization of the tensor modes in this study.

We consider an anisotropic inflation model we proposed in \cite{Watanabe:2009ct}. 
The model includes a vector field coupled to the inflaton field $\phi$ through a kinetic term of the form ${\cal L}_{\rm vec} = -1/4f(\phi) ^2 F_{\mu\nu}F^{\mu\nu}$. 
 The anisotropic inflation predicts the following modification
  to primordial power spectra \citep{Dulaney:2010sq,Gumrukcuoglu:2010yc,Watanabe:2010fh}: 
\[
{\rm i})~ P^{00}({\bf k}) = P_{\rm s}(k) [1+g \sin ^2 \theta ],
\]
\[
{\rm ii})~ P_t ^{\rm unp}({\bf k}) =  P_{\rm t}(k) [1+g_h \sin ^2 \theta ], 
\]
\begin{equation}
{\rm iii})~ P^{0{\rm d}}({\bf k}) 
=  P^{{\rm d}0}({\bf k}) =\sqrt{P_{\rm s}(k)P_{\rm t}(k)} g_c \sin ^2 \theta, \label{eq:inf}
\end{equation}
\[
{\rm iv})~ P_t ^{\rm pol}({\bf k}) =  P_{\rm t}(k) g_l \sin ^4 \theta , 
\]
where $P_{\rm s}(k), P_{\rm t}(k)$ are isotropic parts of scalar and tensor power spectra, the $\theta$ is the angle between $\hat{\bf k}$ and a certain privileged direction, and $k\equiv |{\bf k}|$. 
For simplicity, here we neglected the scale dependence of $g, g_h, g_l, g_c$ which is not significant for the scales relevant to the CMB. The model predicts the consistency relation $g_h = \frac{1}{4}\epsilon g,\ g_c = \sqrt{\epsilon} g$, where $\epsilon$ is a slow-roll parameter, in addition to the usual relation for tensor-to-scalar ratio $r=2P_{\rm t}(k)/P_{\rm s}(k)=16\epsilon$, and the linear polarization is relatively small $g_l \sim {\cal O}(g_h^2)$ and there is no cross correlation between scalar perturbations and cross mode tensor perturbations $P^{\times 0}=P^{0 \times}=0$.

\section{Angular Power Spectra}
In this section, we evaluate the following angular power spectra:
\begin{eqnarray}
C_{ll'mm'}^{XX'} &\equiv& \langle a_{lm}^{X}a_{l'm'}^{X'*} \rangle, \nonumber\\
a_{lm}^{X}(\eta _o, {\bf x}_o) &=& \int d\Omega _{\hat{\bf p}} X(\hat{\bf p},\eta _o , {\bf x}_o) 
Y_{lm}^{*}(\hat{\bf p};{\bf e})
. \label{eq:sh}
\end{eqnarray}
where $*$ denotes complex conjugate and $X=T,E,B$ designates fluctuations in temperature and polarization modes of the CMB.
The polar angle of spherical harmonics
 $Y_{lm}(\hat{\bf p};{\bf e})$ is measured from the direction ${\bf e}$. 
 In the presence of statistical anisotropy, the spectra $C_{ll'mm'}^{XX'}$
are generally dependent on the direction ${\bf e}$. In this paper, we make this direction coincide with the privileged direction of statistical anisotropy for simplicity.

The fluctuation $X$ can be expressed in terms of
 the primordial fluctuation with wavenumber ${\bf k}$ in the following way:
\begin{eqnarray}
X(\hat{\bf p},\eta _o, {\bf x}_o) &=& \int \frac{d^3{\bf k} }{(2\pi )^3} \sum _{L} \sum _{s=-2}^{2} R^{s}({\bf k}) Y_{Ls}(\hat{\bf p};\hat{\bf k}) \nonumber\\
&\ &\times \Delta _{Ls}^{X}(k,\eta _o) e^{i{\bf k}\cdot {\bf x}_o}. \label{eq:X}
\end{eqnarray}
where $s=0,\pm 2$ denote contributions from 3d-scalar, tensor mode, respectively. The vector modes are hereafter neglected for simplicity. Note that $Y_{Ls}(\hat{\bf p};\hat{\bf k})$ explicitly indicates that the polar angle of the spherical harmonics is measured from the direction $\hat{\bf k}$ while the azimuthal angle is assumed to be defined by the polarization bases of $R^{s}$.   
And the transfer functions $\Delta$ satisfy the relations 
\begin{equation}
\Delta_{L,-s}^{M}= \Delta_{L,s}^{M},\ \Delta_{L,-s}^{E} = \Delta _{L,s}^{E},\ \Delta_{L,-s}^{B} = - \Delta _{L,s}^{B}.\label{eq:transfer}
\end{equation}

Substituting Eq.(\ref{eq:X}) into Eq.(\ref{eq:sh}) and 
using the following formula 
\begin{equation}
Y_{lm}(\hat{\bf p};\hat{\bf k}) = \sqrt{\frac{4\pi}{2l+1}} \sum_{m'} Y_{lm'}(\hat{\bf p};{\bf e}) \ _{-m}Y_{lm'}^{*}(\hat{\bf k};{\bf e}) \ ,
\end{equation}
 we have
\begin{eqnarray}
C_{ll'mm'}^{XX'} &=& \int \frac{k^2dk}{(2\pi) ^6} \sum_{s,s'}\tilde{\Delta}_{l,s}^{X}(k,\eta _o) \tilde{\Delta}_{l',s'}^{X'}(k,\eta _o)\\
&\ &\times \int d\Omega _{\hat{\bf k}} P^{ss'}({\bf k}) 
\ _{-s}Y_{lm}^{*}(\hat{\bf k};{\bf e}) \ _{-s'}Y_{l'm'}(\hat{\bf k};{\bf e}),
\nonumber
\end{eqnarray}
where we defined $\tilde{\Delta}_{l,s}^{X}(k,\eta _o)\equiv \sqrt{\frac{4\pi}{2l+1}} \Delta_{l,s}^{X}(k,\eta _o),$ 
$_0Y_{lm}\equiv Y_{lm}$ and $_{\pm2}Y_{lm}$ denote spin weighted spherical harmonics defined in Appendix A. Hereafter, we omit the direction ${\bf e}$ for simplicity.
Then $P^{ss'}$ is associated to the linear polarizations and scalar-tensor correlation via Eq.(\ref{eq:relation}) and the property of transfer function Eq.(\ref{eq:transfer}) helps to simplify the expression.
We see that in the conventional cosmology the assumption $P^{ss'}({\bf k})=\delta_{ss'} P^{|s|}(k)$ (i.e. statistically isotropic, no cross correlation between scalar and tensor, no circular polarization), together with the relation $\Delta _{l,-s}^{B}=-\Delta_{l,s}^{B}$ results in $C^{TB}_{ll'mm'}=C^{EB}_{ll'mm'}=0$

Now, we can see the imprints of statistical anisotropy on the CMB. 

\subsection{ (i) anisotropy in scalar perturbations}
 We can calculate TT,TE,EE spectra in the following way:
\begin{eqnarray}
C_{ll'mm'}^{XX'{\rm (i)}} &=&  \int \frac{k^2dk}{(2\pi) ^6} \tilde{\Delta}_{l,0}^{X} \tilde{\Delta}_{l',0}^{X'} I^{\rm (i)}_{ll'mm'}, \nonumber\\
I^{\rm (i)}_{ll'mm'} &\equiv & \int d\Omega _{\hat{\bf k}} P^{00}({\bf k})  \ Y_{lm}^{*}(\hat{\bf k}) \ Y_{l'm'}(\hat{\bf k}),
\end{eqnarray}
This can be evaluated by expanding the spectrum into spherical harmonics $P^{00}({\bf k})=\sum_{LM} a_{LM}^{00}(k) Y_{LM}(\hat{\bf k})$ where $a_{LM}^{00}(k) =0$ for odd $L$.
Then using the relation
\begin{eqnarray}
&\ &\int d\Omega \ Y_{LM}\ _{-s}Y_{lm}^{*}\ _{-s}Y_{l'm'} \nonumber\\
&=& \sqrt{ \frac{(2L+1)(2l'+1)}{4\pi(2l+1)}} {\cal C}_{LMl'm'}^{lm} {\cal C}_{L0l's}^{ls},
\end{eqnarray}
we have
\begin{equation}
I^{\rm (i)}_{ll'mm'}=\sum_{LM}a_{LM}^{00} \sqrt{\frac{(2L+1)(2l'+1)}{4\pi (2l+1)}} {\cal C}^{lm}_{LMl'm'}{\cal C}^{l0}_{L0l'0} \ .
\end{equation}
As $L$ is even, this is only non-zero for even $l-l'$. The statistical anisotropy is characterized by the component $a_{20}^{00} = -\frac{4}{3}\sqrt{\frac{\pi}{5}}gP_s(k)$ and hence we have
\begin{equation}
I^{\rm (i)}_{ll'mm'}=-\frac{2}{3}gP_s(k)\sqrt{\frac{2l'+1}{2l+1}}{\cal C}^{lm}_{20l'm'}{\cal C}^{l0}_{20l'0} \ ,
\end{equation}
where ${\cal C}$ denotes Clebsch-Gordan coefficient.
This has $m$ dependent contribution for $l'=l,l\pm2$.
The correlation is proportional to $\delta _{mm'}$ due to the axisymmetry of the system and  the fact that we have chosen a specific coordinate in defining spherical harmonics.
 For a general coordinate system, modes with different $m$ are cross correlated.  
Similar analyses are made in \cite{Ackerman:2007nb, Gumrukcuoglu:2007bx}

\subsection{(ii) anisotropy in tensor perturbations}
The induced spectra are
\begin{eqnarray}
C_{ll'mm'}^{XX'\rm (ii)} &=&  \int \frac{k^2dk}{(2\pi) ^6} \tilde{\Delta}_{l,2}^{X} \tilde{\Delta}_{l',2}^{X'} I^{\rm (ii) \pm }_{ll'mm'}, \nonumber\\
I^{\rm (ii)\pm}_{ll'mm'} &=&\int d\Omega _{\hat{\bf k}} P_t^{\rm unp}({\bf k})  \Big( \ _{-2}Y_{lm}^{*}(\hat{\bf k}) \ _{-2}Y_{l'm'}(\hat{\bf k})\nonumber\\
&\ & \quad \pm \ _{+2}Y_{lm}^{*}(\hat{\bf k}) \ _{+2}Y_{l'm'}(\hat{\bf k}) \Big) \ ,
\end{eqnarray}
where the upper / lower sign appears in $TT,EE,TE,BB$ / $TB,EB$ correlations.
We decompose it as: $\ P^{\rm unp}({\bf k})=\sum_{LM} a_{LM}^{\rm unp}(k) Y_{LM}(\hat{\bf k})$ where $a_{LM}^{\rm unp}(k)=0$ for odd $L$. Then, we have
\begin{eqnarray}
I^{\rm (ii)\pm}_{ll'mm'} &=& \sum_{LM}a_{LM}^{\rm unp} \sqrt{\frac{(2L+1)(2l'+1)}{4\pi (2l+1)}}\nonumber\\
&\ &\quad \times{\cal C}^{lm}_{LMl'm'} \left( {\cal C}^{l2}_{L0l'2} \pm  {\cal C}^{l(-2)}_{L0l'(-2)} \right) \ .
\end{eqnarray}
From the symmetry of Clebsch-Gordan coefficient ${\cal C}^{c\gamma}_{a\alpha b\beta}=(-1)^{a+b-c}{\cal C}^{c-\gamma}_{a-\alpha b-\beta}$, we have following relations:
$I^{\rm (ii)+}_{ll'mm'} =0$ for odd $l-l'$, and $I^{\rm (ii)-}_{ll'mm'} =0$ for even $l-l'$. This is a manifestation of parity symmetry of the system. In the case of anisotropic inflation, we have a component $a^{\rm unp}_{20}=-\frac{4}{3}\sqrt{\frac{\pi}{5}}g_hP_t(k)$, which causes non-zero $TT,EE,BB,TE$ correlations for $l'=l,l\pm 2$ and $TB,EB$ correlations for $l'=l\pm 1$.

\subsection{(iii) cross correlation}
We have TT,EE,TE spectra induced by the cross correlation of scalar perturbations and plus mode tensor perturbations:
\begin{eqnarray}
C^{XX'\rm (iii)}_{ll'mm'} &=& \frac{1}{\sqrt{2}}
\int \frac{k^2dk}{(2\pi )^6} 
\left[\tilde{\Delta}_{l0}^{X} \tilde{\Delta}_{l'2}^{X'} I^{\rm (iii)+}_{ll'mm'} \right.\nonumber\\
&\ & \left.
+ (-1)^{l+l'+m+m'}\tilde{\Delta}_{l2}^{X} \tilde{\Delta}_{l'0}^{X'}
I^{\rm (iii)+}_{l',l,-m',-m}
\right],
\end{eqnarray}
and TB,EB spectra:
\begin{equation}
C^{XB\rm (iii)}_{ll'mm'} = \frac{1}{\sqrt{2}}
\int \frac{k^2dk}{(2\pi )^6} \tilde{\Delta}_{l0}^{X} \tilde{\Delta}_{l'2}^{B} I^{\rm (iii)-}_{ll'mm'} \ .
\end{equation}
 Here, we have defined
\begin{eqnarray}
I^{\rm (iii) \pm}_{ll'mm'} &=& \int d\Omega _{\hat{\bf k}} P^{0 \rm d}({\bf k})  \ Y_{lm}^{*}(\hat{\bf k}) \nonumber\\
&\ & \times \left( _{-2}Y_{l'm'}(\hat{\bf k}) \pm _{+2}Y_{l'm'}(\hat{\bf k}) \right) \ ,
\end{eqnarray}
To derive these relations we used the property $P^{d0}({\bf k})=P^{0d}({\bf -k})$ and $ _s Y_{lm}^{*}({\bf k})=(-1)^{l+m}\ _{-s}Y_{l,-m}({\bf -k})$.

The direction dependence of cross correlation is $\sin ^2 \theta$. Hence, TT,EE,TE spectra
 can be evaluated as:
\begin{eqnarray}
I^{\rm (iii)+}_{ll'mm'} &=& 2\Big( \alpha^{-2}_{l+2,m}\delta_{l',l+2}  +\alpha ^{0}_{l,m}\delta_{l',l}  + \alpha ^{+2}_{l-2,m} \delta_{l',l-2}  \Big) \nonumber\\
&\ &\ \times \delta_{mm'} g_c \sqrt{P_s(k)P_t(k)}, \nonumber\\
I^{\rm (iii)-}_{ll'mm'} &=& 2\Big(\beta^{-1}_{l+1,m} \delta_{l',l+1}  + \beta ^{+1}_{l-1,m}\delta_{l',l-1}  \Big) \nonumber\\
&\ &\ \times \delta_{mm'} g_c \sqrt{P_s(k)P_t(k)} \,,\label{eq:iiid}
\end{eqnarray}
where the coefficients are given by:
\[
\alpha ^{+2}_{l,m} \equiv  \sqrt{\frac{l(l-1)(l+m+1)(l-m+1)(l+m+2)(l-m+2)}{(l+1)(l+2)(2l+1)(2l+3)^2(2l+5)}}\,,
\]
\[
\alpha ^{-2}_{l,m} \equiv \sqrt{\frac{(l+1)(l+2)(l+m)(l-m)(l+m-1)(l-m-1)}{(l-1)l(2l-3)(2l-1)^2(2l+1)}}\,,
\]
\[
\alpha ^{0}_{l,m} \equiv \frac{2\left\{3m^2 -l(l+1)\right\}  }{(2l-1)(2l+3)}
\sqrt{\frac{(l-1)(l+2)}{l(l+1)}}\,,
\]
\[
\beta ^{+1}_{l,m} \equiv 2m\sqrt{\frac{(l-1)(l+m+1)(l-m+1)}{l(l+1)(l+2)(2l+1)(2l+3)}}
\,, 
\]
\begin{equation}
\beta ^{-1}_{l,m} \equiv -2m \sqrt{\frac{(l+2)(l+m)(l-m)}{(l-1)l(l+1)(2l+1)(2l-1)}} \,.\label{eq:coe}
\end{equation}
The derivation is given in Appendix \ref{sc:anisotropic}.

\subsection{(iv)linear polarization}
For this case, we have
\begin{eqnarray}
C_{ll'mm'}^{XX'{\rm (iv)}} &=&  \int \frac{k^2dk}{(2\pi) ^6} \tilde{\Delta}_{l,2}^{X} \tilde{\Delta}_{l',2}^{X'} I^{\rm (iv)\pm}_{ll'mm'}, \nonumber\\
I^{\rm (iv)\pm}_{ll'mm'} &=& \int d\Omega _{\hat{\bf k}} P_t^{\rm pol}({\bf k})  \ \Big( \ _{-2}Y_{lm}^{*}(\hat{\bf k})  _{+2}Y_{l'm'}(\hat{\bf k}) \nonumber\\
&\ &\quad \pm _{+2}Y_{lm}^{*}(\hat{\bf k})  _{-2}Y_{l'm'}(\hat{\bf k}) \Big) \ ,
\end{eqnarray}
where upper and lower sign correspond to TT, EE, BB, TE and TB, EB spectra, respectively.
 In the case of anisotropic inflation, where direction dependence is proportional to $\sin ^4 \theta$ we can evaluate them as:
\begin{eqnarray}
&&\frac{1}{2}I^{\rm (iv) +}_{ll'mm'} \nonumber\\
&&= \Big[ \alpha ^{+2}_{lm} \alpha ^{-2}_{l+4,m} \delta _{l',l+4} \nonumber\\
&& + \left(\alpha ^{+2}_{lm} \alpha ^0_{l+2,m}+ \alpha ^0_{l,m}\alpha ^{-2}_{l+2,m} - \beta ^{+1} _{lm}\beta ^{-1}_{l+2,m} \right) \delta _{l',l+2} \nonumber\\
&& + \left( (\alpha^{+2}_{lm})^2+(\alpha^{0}_{lm})^2+(\alpha^{-2}_{lm})^2-(\beta^{-1}_{lm})^2-(\beta^{+1}_{lm})^2 \right) \delta_{l'l} \nonumber\\
&& + \left( \alpha ^0_{lm}\alpha^{+2}_{l-2,m}+\alpha ^{-2}_{lm}\alpha^{0}_{l-2,m}-\beta ^{-1}_{lm}\beta^{+1}_{l-2,m} \right) \delta _{l',l-2} \nonumber\\
&& + \alpha ^{-2}_{lm} \alpha ^{+2}_{l-4,m} \delta_{l',l-4} \Big] \delta _{mm'}g_lP_t(k) 
\end{eqnarray}
and
\begin{eqnarray}
&& \frac{1}{2}I^{\rm (iv)-}_{ll'mm'} \nonumber\\
&&= \Big[ (\beta ^{+1}_{lm}\alpha ^{-2}_{l+3,m}-\alpha ^{+2}_{lm} \beta ^{-1}_{l+3,m})\delta _{l',l+3} \nonumber\\
&&+ ( \beta^{+1}_{lm} \alpha^{0}_{l+1,m} + \beta ^{-1}_{lm} \alpha ^{-2}_{l+1,m}-\alpha ^{+2}_{lm} \beta ^{+1}_{l+1,m} \nonumber\\
&\ &\quad \quad 
-\alpha ^{0}_{lm} \beta ^{-1}_{l+1,m} ) \delta _{l',l+1} \nonumber\\
&& +( \beta^{+1}_{lm} \alpha^{+2}_{l-1,m} + \beta ^{-1}_{lm} \alpha ^{0}_{l-1,m}-\alpha ^{0}_{lm} \beta ^{+1}_{l-1,m} \nonumber\\
&\ &\quad \quad 
-\alpha ^{-2}_{lm} \beta ^{-1}_{l-1,m} ) \delta _{l',l-1} \nonumber\\
&&+ (\beta ^{-1}_{lm}\alpha ^{+2}_{l-3,m}-\alpha ^{-2}_{lm} \beta ^{+1}_{l-3,m})\delta _{l',l-3} \Big] \delta _{mm'}g_lP_t(k). \label{eq:iv}
\end{eqnarray}

\begin{figure}
\includegraphics[width=84mm]{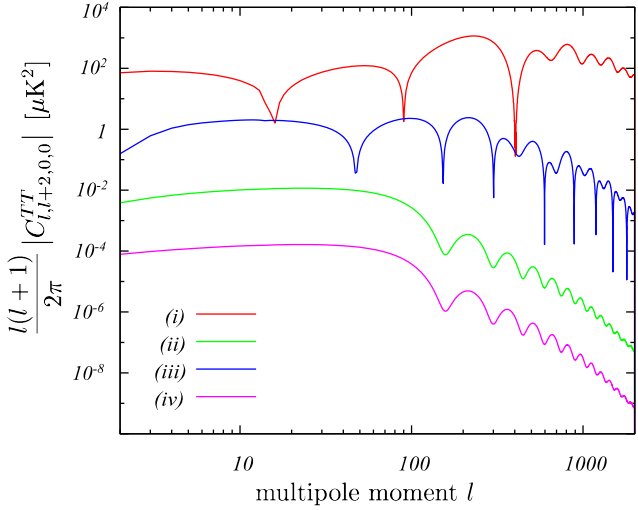}
 \caption{The TT spectra induced by (i) anisotropy in scalar perturbations, (ii) that in tensor perturbations, (iii) cross correlation, and (iv) linear polarization of tensor perturbations. The parameters are chosen as $g=0.3,\ r=0.3$.}
 \label{fg:temp}
\end{figure}
\begin{figure}
\includegraphics[width=84mm]{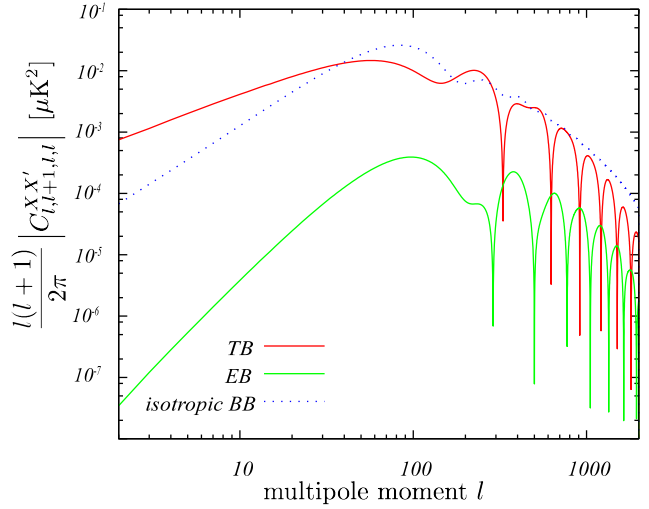}
 \caption{The TB and EB spectra induced by (iii) cross correlation. As a reference of magnitude, the conventional BB spectrum $l(l+1)C_{llmm}^{BB}/2\pi$ induced by isotropic part of the tensor perturbations is plotted with a dotted line. The parameters are chosen as $g=0.3,\ r=0.3$.}
 \label{fg:tbeb}
\end{figure}

\section{discussion}
In the previous section, we have shown that the anisotropy related to tensor perturbations generally induces off-diagonal $TB,EB$ spectra as well as on- and off-diagonal $TT,EE,BB,TE$ spectra. Here we discuss its significance. 

First, we compare the amplitudes of signals induced by the four components of anisotropy.
In Fig \ref{fg:temp}, we have depicted contributions of each component to an off-diagonal $TT$ correlation $\frac{l(l+1)}{2\pi}|C^{TT}_{l,l+2,0,0}|$. 
As for the parameter of anisotropy in scalar perturbations, we adopted the value $g=0.3$
as a reference, which is just of the order of a systematic error in WMAP 5-year data obtained in \cite{Groeneboom:2009cb}. Note that, according to \cite{Pullen:2007tu}, a signal as small as 2\% can be detected with the PLANCK. 
We also assumed the tensor-to-scalar ratio to be $r = 0.3$. Then the other quantities can be determined by the consistency relations in our model: $r=16\epsilon ,\ g_h=\frac{1}{4}\epsilon g,\ g_c=\sqrt{\epsilon} g,\ g_l \sim g_h^2$. 

We see that the contributions of tensor perturbations (ii),(iii) and (iv) are suppressed in comparison to that of (i). And, the cross correlation (iii) has the largest contribution next to (i). This reflects the hierarchy among $rg_h ={\cal O}(g \epsilon ^2), \sqrt{r}g_c = {\cal O}(g \epsilon), rg_l ={\cal O}(g^2 \epsilon ^3)$.
 It is also true for $EE$ and $TE$ spectra.
The ratio between (i), (ii) and (iii) are given by the slow-roll parameter $\epsilon$ (or tensor-to-scalar ratio $r$) and does not depend on the value of $g$. 

Next we consider peculiar signals of anisotropic inflation.
The components (ii), (iii), (iv) induce $B$ mode polarization, and the largest correlation is produced by the cross correlation (iii).
In Fig \ref{fg:tbeb}, we have depicted examples of $TB$ and $EB$ correlations $\frac{l(l+1)}{2\pi}|C^{TB}_{l,l+1,l,l}|, \frac{l(l+1)}{2\pi}|C^{EB}_{l,l+1,l,l}|$. The parameters are again $r=0.3, g=0.3$. As a reference, the conventional $BB$ spectrum induced by the isotropic part of tensor perturbations $\frac{l(l+1)}{2\pi}C^{BB}_{llmm}$ (independent of $m$) is also plotted with a dotted line.
Note that unlike parity violating cases for which odd parity correlations $C_{llmm}^{TB}, C_{llmm}^{EB}$ exist \citep{Lue:1998mq,Liu:2006uh}, our model predicts even parity correlations such as $C_{l,l+1,m,m}^{TB}$ as the result of parity symmetry of the system.
It should be emphasized that neither of the figures contains contribution from monopole components as we depicted off-diagonal correlations ($l\neq l'$).

The ratio of $TB$ correlation induced by cross correlation to the isotropic $BB$ correlation is not dependent on $\epsilon$ (or $r$) for a fixed value of $g$ in our anisotropic inflation model. For the optimistic value of $g \sim 0.3$,
 both amplitudes become comparable. This simple order estimation implies 
 that the $TB$ signal could be comparable to that 
of $B$ mode correlation induced by primordial gravitational wave.
Hence, anisotropic inflation can be a potential source of off-diagonal $TB$ correlation, in addition to other effects such as gravitational lensing and our peculiar velocity proposed so far \citep{Okamoto:2003zw, Amendola:2010ty}.

The correlation induced by  the linear polarization of tensor perturbations (iv) is highly suppressed and lacks any distinctive signature in contrast to circular polarization of tensor perturbations, which predicts odd-parity correlations \citep{Lue:1998mq,Saito:2007kt} in the CMB.

The predicted correlations are $m$ dependent (coordinate dependent) and hence simple summation over $m$ doesn't make sense in constructing the observables which presents the characteristics of statistical anisotropy. Instead bipolar power spectrum \citep{Hajian:2003qq} can be extended to include these signature and used as a spectroscopic tool to distinguish it from other origins of (apparent) statistical anisotropy. 

\section*{Acknowledgments}

MW is supported by JSPS Grant-in-Aid for Scientific Research No. 22E926.
SK is supported in part by grant PHY-0855447 from the National Science 
Foundation.
JS is supported by  the
Grant-in-Aid for  Scientific Research Fund of the Ministry of 
Education, Science and Culture of Japan No.22540274, the Grant-in-Aid
for Scientific Research (A) (No. 22244030), the
Grant-in-Aid for  Scientific Research on Innovative Area No.21111006
and the Grant-in-Aid for the Global COE Program 
``The Next Generation of Physics, Spun from Universality and Emergence".

\bibliography{IAICv2ref.bib}

\appendix

\section{Definition of $\alpha_{lm}^{\pm 2,0},\beta_{lm}^{\pm 1}$}
\label{sc:anisotropic}
In this appendix we present a tool for evaluating $I^{\rm(iii)\pm}_{ll'mm'}$ and $I^{\rm(iv)\pm}_{ll'mm'}$ in a case of anisotropic inflation, where the direction dependence in cross correlation and linear polarization are given by $P^{0\rm d} \propto \sin ^2 \theta,\ P^{\rm pol}_t \propto \sin ^4 \theta $.
First, spin weighted spherical harmonics are associated to spherical harmonics as: 
\begin{eqnarray}
\ _{\pm 2}Y_{lm} &=& (\hat{A}\pm i \hat{B})Y_{lm} \nonumber\\
\hat{A} &=& \frac{- \cos \theta \sin \theta \partial _\theta +\sin ^2 \theta \partial _\theta ^2-\partial _\phi ^2}{\sin^2 \theta \sqrt{(l-1)l(l+1)(l+2)}} 
\,,\nonumber\\
\hat{B} &=& \frac{2(\sin \theta \partial _\theta - \cos \theta )\partial _\phi}{\sin ^2 \theta \sqrt{(l-1)l(l+1)(l+2)}} 
\,.
\end{eqnarray}

Using $\partial _\phi Y_{lm} = imY_{lm}$ and the recursion relations:
\begin{eqnarray}
\sin \theta \partial _\theta Y_{lm} &=& l\sqrt{\frac{(l+1)^2-m^2}{(2l+1)(2l+3)}} Y_{l+1,m}\nonumber\\
&\ &-(l+1)\sqrt{\frac{l^2-m^2}{(2l+1)(2l-1)}}Y_{l-1,m },\nonumber\\
\cos \theta Y_{lm} &=& \sqrt{\frac{(l+1)^2-m^2}{(2l+1)(2l+3)}} Y_{l+1,m}\nonumber\\
&\ & + \sqrt{\frac{l^2-m^2}{(2l+1)(2l-1)}}Y_{l-1,m }\nonumber \ ,
\end{eqnarray}
we obtain the relations:
\begin{eqnarray}
\sin ^2 \theta \hat{A} Y_{lm} &=& \alpha ^{+2}_{lm} Y_{l+2,m} 
+ \alpha ^0_{lm} Y_{lm} + \alpha ^{-2}_{lm}Y_{l-2,m}, 
\nonumber\\
\sin ^2 \theta \hat{B} Y_{lm} &=& i \beta ^{+1}_{lm} Y_{l+1,m} + i\beta ^{-1}_{lm} Y_{l-1,m}.
\end{eqnarray}
The coefficients $\alpha,\ \beta$ are given in Eq.(\ref{eq:coe}). 
Then the orthonormality of spherical harmonics leads to Eq.(\ref{eq:iiid}) and Eq.(\ref{eq:iv}).

\bsp

\label{lastpage}

\end{document}